\documentclass[preprint,1p,12pt]{elsarticle}
\usepackage{soul,hyperref}
\usepackage{graphicx}
\usepackage[latin1]{inputenc}
\usepackage{amsmath,amssymb,amsfonts}

\usepackage{color}
\usepackage{multirow,array}%multicol
\usepackage [ all ]{xy}
\usepackage{pict2e,picture}
\usepackage{epsfig}
\usepackage{xcolor,tikz}
\usetikzlibrary{fit}
\usepackage{verbatim}
\usepackage{nicefrac}
\usepackage[normalem]{ulem}
\usepackage{comment}

\usepackage[ruled,vlined,linesnumbered]{algorithm2e}
\newcommand{\ext}{\mathcal E}
\newcommand{\inte}{\mathcal I}

\renewcommand{\setminus}{\backslash}

\newtheorem{theorem}{Theorem}
\newtheorem{corollary}[theorem]{Corollary}
\newtheorem{lemma}[theorem]{Lemma}
\newtheorem{proposition}[theorem]{Proposition}
\newdefinition{definition}[theorem]{Definition }
\newdefinition{remark}[theorem]{Remark }
\newdefinition{example}[theorem]{Example }
\newproof{proof}{Proof}

\newcommand{\Obg}{\text{Obg}}
\newcommand{\Atg}{\text{Atg}}

\newcommand{\DM}{\text{DM}}

	\definecolor{naranjauca}{cmyk}{ 0, 0.6, 1, 0}
\newcommand{\card}{\text{card}}

\journal{J. Computational and Applied Mathematics}
\date{September 10, 2020}
\makeatletter
\def\ps@pprintTitle{%
  \let\@oddhead\@empty
  \let\@evenhead\@empty
  \let\@evenfoot\@oddfoot
}
\makeatother

\begin{document}
	
	\pagenumbering{roman}
	\medskip
	
	\newpage
	\pagenumbering{arabic}
	
	\begin{frontmatter}
		
		\title{ Impact of local congruences in\\ variable selection from datasets}
		
		\author%[uca]
		{Roberto G. Arag\'on, Jes\'{u}s Medina, Elo\'isa Ram\'irez-Poussa}

		\address%[uca]
		{Department of Mathematics,
			University of  C\'adiz. Spain\\
			Email: \texttt{\{roberto.aragon,jesus.medina,eloisa.ramirez\}@uca.es}}

		\begin{abstract}
Formal concept analysis (FCA) is a useful mathematical tool for obtaining information from relational datasets. One of the most interesting research goals in FCA is the selection of the most representative variables of the dataset, which is called attribute reduction. Recently, the attribute reduction  mechanism has been complemented with the use of local congruences in order to obtain robust clusters of concepts, which form convex sublattices of the original concept lattice. Since the application of such local congruences modifies the quotient set associated with the attribute reduction, it is fundamental to know how the original context (attributes, objects and relationship) has been modified in order to understand the impact of the application of the local congruence in the attribute reduction. 
 
		\end{abstract}

	\end{frontmatter}

\section{Introduction}
The variable selection problem is a hot topic in many areas dedicated to data analysis. On many occasions, making a correct selection of the considered variables facilitates the management of the information in a considerable way, but it can also lead to certain changes in the provided information that must be analyzed in order to control them. This is one of the most appealing research lines within the theory of formal concept analysis.

Formal Concept Analysis (FCA){~\cite{GanterW}} is a mathematical theory to organize and analyze the information collected in a dataset, by means of the mathematical structure called concept lattice. Since its introduction~\cite{GanterW}, several mechanisms for variable selection have been intensively studied. One of the most researched lines deals with the reduction of the number of attributes, detecting the unnecessary ones {and} preserving the most important information of the considered formal context~\cite{buruscoipmu2018,Antoni2016,chen2019,ar:ins:2015,Cornejo2017,ins2018:cmr,konecny2019,camwa-medina,Ren2016,Shao2017}.

In~\cite{bmrd:RSTFCA:c,bmrd:FCARST:f}, {the} authors proved that any attribute reduction of a formal context induces an equivalence relation on the set of concepts of the concept lattice. Moreover, the equivalence classes of the induced equivalence relation have the structure of a join-semilattice. This fact gave rise to the introduction  of the definition of a new equivalence relation whose main goal was to {lightly modify} the  equivalence relation {induced by the attribute reduction}, {in order to provide}  new {more robust} equivalence classes {grounded on}  the structure of convex sublattices. This new equivalence relation was called local congruence and the application to attribute reduction {in FCA} was initiated in~\cite{ aragonIJCRS20,aragonSCI20}.

In addition,\cite{aragon:ipmu:2020} studied  that when the induced equivalence relation provided by an attribute reduction does not coincide with a local congruence, then the fact of using a local congruence to complement such a reduction had an influence on the original reduction.

 In this paper, we continue with the idea {highlighted  in}~\cite{aragon:ipmu:2020} of studying the impact of local congruences on concept lattices corresponding to formal contexts that have been previously reduced {deleting a set of (unnecessary) attributes (attribute reduction)}. {For that,} a new partial order{ing on the quotient set provided by a local congruence is introduced and analyzed. Moreover, the structures associated with the reduced context and the quotient set of concepts given by an attribute reduction have been compared in order to highlight the narrow relationship between them, which will be fundamental for simplifying and proving the rest of results of the paper.}

{Since the main transformation that a local congruence  makes on the quotient set associated with the attribute reduction is to group different classes, this paper studies  how the reduced context (attributes, objects and the relationship) needs to be modified (as less as possible) in order to obtain a complete lattice isomorphic to the one obtained after the grouping  given by the local congruence. This study will be split into three different cases depending on the character of the element grouped by the local congruence, specifically, if it is join-irreducible, meet-irreducible or neither join nor meet-irreducible. 
Furthermore, due to the modification of the original context  from the attribute reduction mechanism is well known, we can assert that this paper determines the precise modifications of an attribute reduction mechanism complemented by a local congruence makes on the original context. Hence, this study is fundamental for the traceability and  knowledge of  the  information obtained  from  the proposed methodology for variable selection from datasets.}

 The paper is organized as follows: Section~\ref{Preli} recalls some necessary notions and results needed in the development of  the contributions of the {paper}. In Section~\ref{imcon:atrired}, we carried out the {relationship between the reduced concept lattice and the quotient set associated with the attribute reduction. Moreover, it is defined and studied an ordering defined on the quotient set associated with a  local congruence.} The results obtained in Section~\ref{imcon:atrired} have led us to develop the analysis shown in the Section~\ref{imrem:elemconcep}, which is divided in two parts: the first one devoted to study the impact of eliminating a join-irreducible element of a concept lattice and the second one devoted to analyze the repercussion of eliminating other kind of elements. Section~\ref{conclusion} summarizes {the} conclusions and prospect for future works.

\section{Preliminaries}\label{Preli}

In this section, 
some preliminary notions and results used in this work will be recalled. In order to make this paper as self-contained as possible, the preliminary section is divided into three parts, the first one will be devoted to recall those necessary notions of FCA, the second one to those related to lattice theory and the last one to local congruences.

\subsection{Formal concept analysis}

In FCA a context is a triple  $(A, B, R)$ where $A$ is a set of attributes, $B$ is a set of objects and 
 {$R\subseteq A\times B$}
  is a relationship, such that {$(a,x)\in R$ (also denoted  as $aRx$)}, if the object $x\in B$ possesses the attribute $a\in A$. In addition, we call \emph{derivation operators} to the  mappings ${\ }^\uparrow\colon 2^B\to 2^A$ and ${\ }^\downarrow\colon 2^A\to 2^B$ defined for each $X\subseteq B$ and $Y\subseteq A$  as:
	\begin{eqnarray}\label{def-classical.gc1}
	X^{\uparrow}&=&\{a\in A\mid \hbox{for all } x\in X, aRx\}\\\label{def-classical.gc2}
	Y^{\downarrow}&=&\{x\in B\mid \hbox{for all } a\in Y, aRx\}
	\end{eqnarray}
Taking into account the previous mappings, a \emph{concept} is a pair $(X,Y)$, with $X \subseteq B$ and $Y \subseteq A$ satisfying that $X^{\uparrow}=Y$ and $Y^\downarrow=X$.  The subset $X$ is called the \emph{extent} of the concept and the subset $Y$ is called the \emph{intent}. The set of extents and intents are denoted by  $\ext(A,B,R)$ and $\inte(A,B,R)$, respectively. 

{The whole set of concepts is denoted as $\mathcal{C}(A,B,R)$. The inclusion ordering on the left argument,
 $\leq$, provides $\mathcal{C}(A,B,R)$ with the structure of a complete lattice, which is called \emph{concept lattice} of the context $(A,B,R)$.}
	
Furthermore, we need to recall the notion of meet(join)-irreducible element of a lattice.

	\begin{definition}\label{def:irred}
Given a lattice $(L,\preceq)$, such that $\wedge$ is the meet operator,  and an element $x\in L$  verifying
\begin{enumerate}
\item If $L$ has a top element $\top$, then $x\neq \top$.
\item If  $x= y\wedge z$, then $x=y$ or $x=z$, for all $y,z\in L$.
\end{enumerate}
we call $x$ \emph{meet-irreducible ($\wedge$-irreducible) element} of $L$.  Condition $(2)$ is equivalent to
\begin{enumerate}
\item[$2'$.]  If $x<y$ and $x<z$, then $x<y\wedge z$,  for all $y,z\in L$.
\end{enumerate}

The \emph{join-irreducible ($\vee$-irreducible) element} of $L$ is defined dually.
	\end{definition}

In addition, we will say that an \emph{attribute-concept} is a concept generated by an attribute $a \in A$, that is $(a^{\downarrow}, a^{\downarrow\uparrow})$. Dually, an \emph{object-concept} is defined as $(b^{\uparrow\downarrow}, b^{\uparrow})$ for $b\in B$. Moreover,  the sets of objects and attributes that generates a concept are defined.

\begin{definition}\label{obg}
	Given a formal context $(A, B, R)$, the associated concept lattice $\mathcal{C}(A,B, R)$ and a concept $C\in \mathcal{C}(A, B, R)$, the set of \emph{objects generating} $C$ is defined as the set:
	$$\Obg(C) = \{b\in B \mid (b^{\uparrow\downarrow}, b^{\uparrow}) = C  \}$$
	Similarly, the set of \emph{attributes generating} $C$ is defined as the set:
	$$\Atg(C) = \{a\in A \mid (a^{\downarrow}, a^{\downarrow\uparrow}) = C  \}$$
\end{definition}

In addition, the sets $\Obg(C)$ and $\Atg(C)$ are always nonempty sets, for every join-irreducible and meet-irreducible concept $C$~\cite{ins2018:cmr}, respectively.
\begin{proposition}\label{prop:atgnonempty}
If $C$ is a join-irreducible concept of $\mathcal{C}(A,B, R)$, then  $\Obg(C)$ is a nonempty set. Equivalently, if $C$ is a meet-irreducible concept of $\mathcal{C}(A,B, R)$, then  $\Atg(C)$ is a nonempty set.
\end{proposition}
{\begin{proof}
 The result straightforwardly arises from  definition.

\end{proof}}

With respect to the reduction of the context in FCA, from the perspective of the set of objects, the classification of the objects based on {the sets given in} Definition~\ref{obg}, is given below. This result is dual to the one given for the set of attributes in~\cite{ins2018:cmr}. For a more detailed information about the notions considered in the this result we refer the reader to~\cite{ins2018:cmr}.

\begin{theorem}\label{th:classificacion1:atgC}
Given an object $b\in B$, {we have that} 
\begin{itemize}
\item $b$ is an absolutely necessary object if and only if there exists a join-irreducible concept $C$ of $(\mathcal{M},\preceq)$, satisfying that $b\in \Obg(C)$ and $\card(\Obg(C))=1$. 
\item $b$ is a relatively necessary object if and only if $b$ is not an absolutely necessary object and there exists  a join-irreducible concept $C$ with $b\in \Obg(C)$ and  $\card(\Obg(C))>1$, satisfying that $\big(B\setminus \Obg(C)\big)\cup\{B\}$ is a consistent set. 
\item $b$ is an absolutely unnecessary object if and only if, for any join-irreducible concept $C$, $b\notin  \Obg(C)$, or if $b\in \Obg(C)$ then $\big(B\setminus \Obg(C)\big)\cup\{b\}$ is not a consistent set.
\end{itemize}
\end{theorem}

In addition, it is important to recall that when we reduce the set of attributes in a context,  an equivalence relation on the set of concepts of the original concept lattice is induced. The following proposition was proved in~\cite{bmrd:RSTFCA:c}  for the classical setting of FCA. 
	\begin{proposition}[\cite{bmrd:RSTFCA:c}]\label{prop:clase2}
		Given a context $(A, B, R)$ and a subset $D \subseteq {A}$.	
		The set $ \rho_D=\{((X_1,Y_1),(X_2,Y_2)) \mid (X_1,Y_1),(X_2,Y_2)\in \mathcal C(A,B, R), X_1^{\uparrow_D\downarrow}= X_2^{\uparrow_D\downarrow}\}$ is an equivalence relation{, w}here ${}^{\uparrow_D}$ denotes the concept-forming operator 
		${X^{\uparrow_D} = \{a\in D\mid (a,x)\in R{, \hbox{for all } x\in X}\}}$
		restricted to the subset of attributes $D\subseteq A$.
	\end{proposition}

{Notice that  $^{\uparrow_D}$ and $^{\downarrow^D}$, respectively defined as above and as $Y^{\downarrow^D} = \{b\in B\mid (y,b)\in R, \hbox{for all } y\in Y\}$, for all $Y\subseteq D$, are  the  concept-forming operators of the reduced concept lattice. Notice that,  since the set of object $B$ is not modified, we have that $Y^{\downarrow^D} =Y^{\downarrow} $, for all $Y\subseteq D$. }

In~\cite{bmrd:RSTFCA:c},  the authors also proved that each equivalence class of the induced equivalence relation has a structure of join semilattice and they also characterized the maximum element. 
	\begin{proposition}[{\cite{bmrd:RSTFCA:c}}]\label{prop:clase3}
		Given a context $(A, B, R)$, a subset $D \subseteq {A}$ and a class $[(X,Y)]_D$ of the quotient set $ \mathcal C(A,B, R)/\rho_D$. The class $[(X,Y)]_D$  is a join semilattice with maximum element ${(X^{\uparrow_D\downarrow},X^{\uparrow_D\downarrow\uparrow}})$.
	\end{proposition}

\subsection{Lattice theory}

In this paper, we will also make use of some well-known notions of lattice theory which are recalled below. The first notion is about the chain conditions on lattices.

\begin{definition}[\cite{DaveyPriestley}]\label{acc_dcc}
	Let $(L,\preceq)$ be a lattice. $L$ is said to satisfy the \emph{ascending chain condition}, denoted as ACC, if given any sequence $x_1 \preceq x_2 \preceq \dots \preceq x_n \preceq \dots$ of elements of $L$, there exists $k\in\mathbb{N}$ such that $x_k = x_{k+1} = \dots$. The dual of the ascending chain condition is the \emph{descending chain condition}, denoted as DCC.
\end{definition}

The following result relates the chain conditions to the completeness of a lattice.

\begin{theorem}[\cite{DaveyPriestley}]\label{th:2.41DP}
	Let $(L, \preceq)$ be a lattice.
	\begin{enumerate}[(i)]
		\item If $L$ satisfies ACC, then for every non-empty subset $A\subseteq L$ there exists a finite subset $F\subseteq A$ such that $\bigvee A = \bigvee F$.
		\item If $L$ has a bottom element and satisfies ACC, then $L$ is complete.
		\item If $L$ has no infinite chain, then $L$ is complete.
	\end{enumerate}
\end{theorem}

On the other hand,  an ordered set can be embedded in a complete lattice by using the Dedekind-MacNeille completion, which is associated with the Galois connection $({\ }^u, {\ }^l)$, recalled in the following definition.

\begin{definition}[\cite{DaveyPriestley}]\label{DMcompletion}
	Let $(P,\leq)$ be an ordered set. The \emph{Dedekind-MacNeille completion} of $P$ is defined as
	$$ \DM(P) = \{A\subseteq P \mid A^{ul} = A\}$$
	where the mappings ${\ }^u\colon 2^P \rightarrow 2^P$ and ${\ }^l\colon 2^P \rightarrow 2^P$ are defined for a subset $A\subseteq P$ as
	\begin{align*}
	A^u = \{x \in P \mid a \leq x \text{, for all } a\in A\} \\
	A^l = \{x \in P \mid {x \leq a} \text{, for all } a\in A\} 
	\end{align*}
	The ordered set $(\DM(P), \subseteq)$ is a complete lattice.
\end{definition}

In addition, we can use the Dedekind-MacNeille completion to construct a complete lattice from the join-irreducible and meet-irreducible elements of a complete lattice as the following result states.

\begin{theorem}[\cite{DaveyPriestley}]\label{DP:Th7.42}
	Let $(L, \preceq)$ be a lattice with no infinite chains. Then
	$$L \cong \DM(\mathcal{J}(L)\cup\mathcal{M}(L))$$
	where $\mathcal{J}(L)$ and $\mathcal{M}(L)$ are the sets of join-irreducible and meet-irreducible elements of $L$, respectively. Moreover, $\mathcal{J}(L)\cup\mathcal{M}(L)$ is the smallest subset of $L$ which is both join-dense and meet-dense in $L$.
\end{theorem}

\subsection{Local congruences}\label{lc}
The notion of local congruence arose with the goal of complementing attribute reduction in FCA. The purpose of local congruences is to obtain equivalence relations less-constraining than congruences~\cite{aragonSCI20} 
and with useful properties to be applied in size reduction processes of concept lattices. We recall the notion of local congruence {next.} 
	
\begin{definition}\label{Def-wc}
	Given a lattice $\left( L, \preceq \right)$, we say that an equivalence relation $\delta$ on $L$ is a \emph{local congruence} if each equivalence class of $\delta$ is a convex sublattice of $L$.
\end{definition}

The notion of local congruence  in terms of the equivalence relation is given as follows.

\begin{proposition}[{\cite{aragonSCI20}}]
 Given a lattice $( L, \preceq)$ and an equivalence relation $\delta$ on $L$, the relation $\delta$ is a local congruence on $L$ if and only if, for each $a, b, c \in L$, the following properties hold:
 \begin{enumerate}
 	\item[(i)] If $(a,b)\in\delta$ and $a\preceq c\preceq b$, then $(a,c)\in\delta$.
 	\item[(ii)] $(a,b)\in\delta$ if and only if $(a\wedge b, a\vee b)\in\delta$.
 \end{enumerate}
 \end{proposition}

Usually, we will look for a local congruence that contains a partition induced by an equivalence relation. When we say that a local congruence contain a partition provided by an equivalence relation, we are making use of the following definition of inclusion ordering of equivalence relations. 

\begin{definition}\label{localinclusion}
	Let $\rho_1$ and $\rho_2$ be two equivalence relations on a lattice $(L,\preceq)$. We say that the equivalence relation \emph{$\rho_1$ is included in $\rho_2$},  denoted as $\rho_1 \sqsubseteq \rho_2$, if for every equivalence class $[x]_{\rho_1} \in L/ \rho_1$ there exists an equivalence class $[y]_{\rho_2} \in L /\rho_2$ such that $[x]_{\rho_1}\subseteq [y]_{\rho_2}$.
	
\end{definition}

Once we have recalled the previous notions and results,  {next section} will investigate the impact of  local congruences on concept lattices when the considered concept lattices are associated with  reduced contexts.

\section{{Attribute reduction quotient sets and local congruences}}\label{imcon:atrired}

{This section will begin showing, in an attribute reduction process, the narrow relationship between   the quotient set associated with the attribute reduction (Proposition~\ref{prop:clase2}) and the  reduced concept lattice.  Then,  an ordering relation will be defined and studied on the quotient set associated with  a local congruence, which will be fundamental for the main goal of this paper.}

{\subsection{Attribute reduction: quotient set versus reduced concept lattice}}

Since the equivalence classes induced by an attribute reduction are join-semilattices with maximum elements, for every equivalence class $[C]_D$, with $C=(X,Y)\in \mathcal C(A,B,R)$, the concept $C_M =(X_M,Y_M)= \underset{C_i\in [C]_D}{\bigvee} C_i$ necessarily belongs to $[C]_D$. Indeed, by Proposition~\ref{prop:clase3}, this maximum element is  ${(X^{\uparrow_D\downarrow},X^{\uparrow_D\downarrow\uparrow}})$ and so,   $X_M =X^{\uparrow_D\downarrow}$, 
which implies that this extent is also the extent of a concept of the reduced concept lattice $\mathcal{C}(D, B, R_{|D\times B})$. 
Moreover, the least element {of   $\mathcal{C}(D, B, R_{|D\times B})$}  is ${(\varnothing^{\uparrow_D\downarrow},\varnothing^{\uparrow_D}})$, which correspond to the concept ${(\varnothing^{\uparrow_D\downarrow},\varnothing^{\uparrow_D\downarrow\uparrow}})$ of the original context. Notice that $\mathcal{C}(D, B, R_{|D\times B})$ is a complete lattice and so, a {join}  closed structure with a least element.

On the other hand, on  the whole set of equivalence classes given by the relation $\rho_D$ an ordering can be defined.
\begin{proposition}
On the quotient set $ \mathcal C(A,B, R)/\rho_D$ associated with a context $(A,B,R)$, the relation $\sqsubseteq_D$, defined as  $[(X_1,Y_1)]_D\sqsubseteq_D[(X_2,Y_2)]_D$ if ${X_1}^{\uparrow_D\downarrow}\subseteq {X_2}^{\uparrow_D\downarrow}$, for all $[(X_1,Y_1)]_D,[(X_2,Y_2)]_D\in  \mathcal C(A,B, R)/\rho_D$, is an ordering relation. 
\end{proposition}
\begin{proof}
By Proposition~\ref{prop:clase3}, $\sqsubseteq_D$ is well defined. Moreover,  from  its definition,  $\sqsubseteq_D$ is straightforwardly reflexive, antisymmetric and transitive. \qed
\end{proof}

The following result shows the narrow relationship between the previously shown quotient set and  the reduced concept lattice, which improves Proposition 3.11 in~\cite{bmrd:RSTFCA:c}.
\begin{theorem}\label{th:isomorph}
 Given a context $(A,B,R)$ and  a subset of attributes $D\subseteq A$, we have that 
 the quotient set given by $\rho_D$ and the reduced concept lattice by $D$ are isomorphic, that is
$$
(\mathcal{C}(A, B, R)/\rho_D,\sqsubseteq_D) \cong (\mathcal{C}(D, B, R_{|D\times B}),\leq_D) 
$$
where $\leq_D$ is the ordering in the original concept lattice restricted to the reduced one. 
\end{theorem}
\begin{proof}
We will define two mappings  $\varphi\colon   \mathcal{C}(A, B, R)/\rho_D \to  \mathcal{C}(D, B, R_{|D\times B}) $ and 
  $\psi\colon   \mathcal{C}(D, B, R_{|D\times B})  \to \mathcal{C}(A, B, R)/\rho_D $, and we will prove that they preserve the ordering and that $\varphi \circ \psi$ and $\psi \circ \varphi$ are the corresponding identity mappings.
  
The mapping  $\varphi$ will be {defined} as   $\varphi([(X,Y)]_D)=(X^{\uparrow_D\downarrow},X^{\uparrow_D})$ for all $[(X,Y)]_D\in \mathcal{C}(A, B, R)/\rho_D$,  and  $\psi$ is {defined} as   $\psi(X,Y) =[(X,X^{\uparrow})]_D$ for all $(X,Y)\in \mathcal{C}(D, B, R_{|D\times B})$.  $\varphi$ is clearly well defined and $\psi$ is also well defined by  Proposition~\ref{prop:clase3}.

Given   $(X,Y)\in \mathcal{C}(D, B, R_{|D\times B})$, we have that
$$
\varphi \circ \psi(X,Y)=\varphi ( \psi(X,Y))=\varphi ([(X,X^{\uparrow})]_D)=(X^{\uparrow_D\downarrow},X^{\uparrow_D}) =(X,Y)
$$
where  the last equality holds because $(X,Y)$ is a concept of the reduced concept lattice and so, it satisfies  $X^{\uparrow_D\downarrow}=X^{\uparrow_D\downarrow^D}=X$.

In the other composition, we consider  $[(X,Y)]_D\in \mathcal{C}(A, B, R)/\rho_D$ and  we obtain
$$
\psi \circ \varphi([(X,Y)]_D)=\psi ( \varphi([(X,Y)]_D))=\psi (X^{\uparrow_D\downarrow},X^{\uparrow_D})=[(X^{\uparrow_D\downarrow},X^{\uparrow_D\downarrow\uparrow})]_D 
$$
and   the last equivalence class is exactly equal to $[(X,Y)]_D$, by  Proposition~\ref{prop:clase3}.

Finally, we will prove that $\varphi$ and $\psi$ are order-preserving. 
Given $[(X_1,Y_1)]_D$, $[(X_2,Y_2)]_D\in \mathcal{C}(A, B, R)/\rho_D$, such that $[(X_1,Y_1)]_D\sqsubseteq_D [(X_2,Y_2)]_D$, we have that ${X_1}^{\uparrow_D\downarrow}\subseteq {X_2}^{\uparrow_D\downarrow}$ by definition of $\sqsubseteq_D$  and consequently we have that
$$
\varphi([(X_1,Y_1)]_D)=({X_1}^{\uparrow_D\downarrow},{X_1}^{\uparrow_D})\leq_D ({X_2}^{\uparrow_D\downarrow},{X_2}^{\uparrow_D}) =\varphi([(X_2,Y_2)]_D)
$$

Now, given $(X_1,Y_1), (X_2,Y_2)\in \mathcal{C}(D, B, R_{|D\times B})$, such that $(X_1,Y_1)\leq_D(X_2,Y_2)$, we obtain that  
$$
\psi(X_1,Y_1) =[(X_1,{X_1}^{\uparrow})]_D \overset{(*)}{\sqsubseteq}_D [(X_2,{X_2}^{\uparrow})]_D=
\psi(X_2,Y_2)
$$
where $(*)$ holds    because ${X_1}^{\uparrow_D\downarrow}{=X_1^{\uparrow_D\downarrow^D}}=X_1\subseteq X_2{=X_2^{\uparrow_D\downarrow^D}}={X_2}^{\uparrow_D\downarrow}$.
\qed
\end{proof}

As a consequence, $(\mathcal{C}(A, B, R)/\rho_D,\sqsubseteq_D)
$ is a complete lattice, which will be taken into account in the relationship with the local congruences.

{\subsection{The poset associated with a local congruence}}
Now, we define a new relationship on the elements of the quotient set provided by a local congruence,  which turns out to be a partial order as we will prove in the following result.

\begin{theorem}\label{newpo}
	Given a complete lattice $(L,\preceq)$ and a local congruence $\delta$ on $L$, the binary relation defined as follows:
	$$ [x]_\delta \leq_\delta [y]_\delta \quad~\text{if }~  \quad
	\bot_L \in [x]_\delta,  \quad\text{or} \quad x_M \preceq y'  
	$$
	where $y'\in [y]_\delta$, $x_M = \bigvee_{x_i\in [x]_\delta} x_i$ and $\bot_L$ is the bottom of $(L,\preceq)$, is a partial order for $L/\delta$.
\end{theorem}
\begin{proof}
	We consider a lattice $(L,\preceq)$, a local congruence $\delta$ on $L$ and the relation $\leq_\delta$ defined above. We are going to prove that the relation $\leq_\delta$ is a reflexive, antisymmetric and transitive relation. It is clear that it is reflexive. In order to prove the antisymmetry property, let us consider two classes $[x]_\delta, [y]_\delta\in L/\delta$ satisfying  that $[x]_\delta\leq_\delta[y]_\delta$ and $[y]_\delta\leq_\delta[x]_\delta$, then  we have to distinguish the following cases:
	\begin{enumerate}
		\item If $\bot_L\in [x]_\delta$ and $\bot_L\in [y]_\delta$, then $[x]_\delta = [y]_\delta$ since the equivalence classes are disjoint.
		\item If $\bot_L\in [x]_\delta$ and $\bot_L\not\in [y]_\delta$, then  {since $[x]_\delta\leq_\delta[y]_\delta$,  we have that there exists $y'\in[y]_\delta$, such that} $\bot_L \preceq y'$. {On the other hand,   since $[y]_\delta\leq_\delta[x]_\delta$,  there exists $x'\in[x]_\delta$, such that   $y_M \preceq x'$. Therefore,}
		$$
		\bot_L \preceq y' \preceq y_M \preceq x'
		$$
		which  implies that $[x]_\delta = [y]_\delta$, by the convexity of the equivalence classes provided by local congruences.  
		\item If $\bot_L\not\in [x]_\delta$ and $\bot_L\not\in [y]_\delta$, then 
		{since $[x]_\delta\leq_\delta[y]_\delta$,  we have that there exists $y'\in[y]_\delta$, such that} $x_M  \preceq y'$. {On the other hand,   since $[y]_\delta\leq_\delta[x]_\delta$,  there exists $x'\in[x]_\delta$, such that   $y_M \preceq x'$. Consequently,}
		$$
		x_M \preceq y' \preceq y_M \preceq x'
		$$
		and, by the convexity of the classes, we can conclude that  $[x]_\delta = [y]_\delta$.
	\end{enumerate}

	Now, we consider three different classes $[x]_\delta, [y]_\delta, [z]_\delta\in L/\delta$ such that $[x]_\delta\leq_\delta[y]_\delta$ and $[y]_\delta\leq_\delta[z]_\delta$ in order to prove the transitivity property. We only have to distinguish the following two cases:
	\begin{enumerate}
		\item If $\bot_L\in [x]_\delta$, $\bot_L\not\in [y]_\delta$ and $\bot_L\not\in [z]_\delta$, then we have straightforwardly that $\bot_L \preceq z'$ for any $z'\in [z]_\delta$ by definition of infimum of the lattice. Therefore, $[x]_\delta \leq_\delta [z]_\delta$.
		\item If $\bot_L\not\in [x]_\delta$, $\bot_L\not\in [y]_\delta$ and $\bot_L\not\in [z]_\delta$, then since $[x]_\delta\leq_\delta[y]_\delta$ and $[y]_\delta\leq_\delta[z]_\delta$, we have that there exist $y'\in [y]_\delta$ and $z'\in[z]_\delta$ such that $x_M \preceq y'$ and $y_M \preceq z'$. Therefore,
			$$x_M \preceq y' \preceq y_M \preceq z',$$
			which implies that $x_M\preceq z'$, i.e. $[x]_\delta \leq [z]_\delta$.
	\end{enumerate}
	Hence, we can conclude that the binary relation $\leq_\delta$ is a partial order.\qed
\end{proof}

{
The condition $x_M \preceq y'$ given in Theorem~\ref{newpo} can clearly be more concrete, i.e., since it is satisfied that $y' \preceq y_M$ for all $y' \in [y]_\delta$, we have that $x_M \preceq y_M$ in particular, is  satisfied. This fact is stated in the following result.

\begin{corollary}
	Given a complete lattice $(L, \preceq)$ and a local congruence $\delta$ on $L$, the partial order defined in Theorem~\ref{newpo} is equivalent to the following:
	$$ [x]_\delta \leq_\delta [y]_\delta \quad~\text{if }~  \quad
	\bot_L \in [x]_\delta  \quad\text{or} \quad x_M \preceq y_M  
	$$
	where $y_M = \bigvee_{y_i\in [y]_\delta} y_i$.
\end{corollary}

}

{Notice that  the previous ordering is associated with the join-sublattice structure given by an attribute reduction (Proposition~\ref{prop:clase3}). {Hence, since we could also include a bottom element, we could think that this ordering can provide a complete lattice.}
However, when we define the quotient set of an {arbitrary} local congruence $\delta$ on a lattice $L$ with the partial order $\leq_\delta$, we do not always obtain that $L/\delta$ is a lattice. 

\begin{example}\label{Example2}
	Let us consider a context $(A, B, R)$ whose relation is given in Table~\ref{tabla2}. We also consider a local congruence $\delta$ on the associated concept lattice {$\mathcal{C}(A, B, R)$} which is shown on the left side of Figure~\ref{Example2:fig1}. 
	
	If we define the quotient set $\mathcal{C}(A, B, R)/\delta$ with the partial order {defined} in Theorem~\ref{newpo}, we obtain that $(\mathcal{C}(A, B, R)/\delta, \leq_\delta)$ is not a lattice but a poset {because the infimum of each pair of elements of $(\mathcal{C}(A, B, R)/\delta, \leq_\delta)$ does not exist, for instance, the lower bounds of the concepts $C_{12}$ and  $C_{13}$ are the concepts $C_3, C_4$ and $C_1$ whence $C_3$ and $C_4$ are minimal, as it can be observed in the right side of Figure~\ref{Example2:fig1}. Hence, the infimum of $C_{12}$ and $C_{13}$ does not exist.}	\qed
	\begin{table}[h]
		\begin{center}
			\begin{tabular}{l|c c c c c c c}
				\hline
				$R$ & $b_1$  & $b_2$ & $b_3$ & $b_4$ & $b_5$ & $b_6$ & $b_7$\\ \hline
				$a_1$ & 1 & 1 & 1 & 1 & 1 & 0 & 0 \\ 
				$a_2$ & 1 & 1 & 0 & 1 & 0 & 0 & 0 \\
				$a_3$ & 1 & 1 & 1 & 0 & 1 & 0 & 0\\
				$a_4$ & 0 & 1 & 1 & 0 & 1 & 1 & 0 \\
				$a_5$ & 0 & 1 & 1 & 0 & 0 & 1 & 1 \\
				$a_6$ & 1 & 0 & 0 & 0 & 0 & 0 & 0 \\
				$a_7$ & 0 & 0 & 1 & 0 & 0 & 0 & 0 \\\hline
			\end{tabular}
		\end{center}
		\caption{Relation of Example~\ref{Example2}.
			\label{tabla2}
		}
	\end{table}
	
	\begin{figure}[h!]
		\begin{minipage}{0.45\textwidth}
			\begin{center}
				\tikzstyle{place}=[circle,draw=black!75, fill=black!75]
				\begin{tikzpicture}[inner sep=0.75mm,scale=0.9, every node/.style={scale=0.8}]				
				\node at (-1,0) (C1) [place, label={[label distance=0.2cm]below:$C_1$}] {};  			
				\node at (-2.5,1) (C2) [place, label={[label distance=0.2cm]below:$C_2$}] {};
				\node at (-1,1) (C3) [place, label={[label distance=0.1cm]left:$C_3$}] {};
				\node at (0.5,1) (C4) [place, label={[label distance=0.2cm]below:$C_4$}] {};
				\node at (-1.9,3) (C5) [place, label={[label distance=0cm]200:$C_5$}] {};
				\node at (-0.1,2) (C6) [place, label={[label distance=0cm]350:$C_6$}] {};
				\node at (-2.5,4) (C7) [place, label={[label distance=0.1cm]left:$C_7$}] {};  			
				\node at (-1.25,4) (C8) [place, label={[label distance=0cm]left:$C_8$}] {};
				\node at (-0.75,3) (C9) [place, label={[label distance=0cm]right:$C_9$}] {};		
				\node at (0.5,3) (C10) [place, label={[label distance=0.1cm]right:$C_{10}$}] {};
				\node at (-1.9,5) (C11) [place, label={[label distance=0.1cm]above:$C_{11}$}] {};
				\node at (-0.1,4) (C12) [place, label={[label distance=0cm]110:$C_{12}$}] {};
				\node at (1,4) (C13) [place, label={[label distance=0.2cm]above:$C_{13}$}] {};
				\node at (-0.5,6) (C14) [place, label={[label distance=0.2cm]above:$C_{14}$}] {};
				
				\draw [-] (C1) -- (C2)-- (C5)-- (C7)-- (C11)-- (C14);
				\draw [-] (C1) -- (C3)-- (C5)-- (C8);
				\draw [-] (C1) -- (C4)-- (C6)-- (C9)-- (C8)-- (C11);
				\draw [-] (C3) -- (C6)-- (C10)-- (C13)-- (C14);
				\draw [-] (C9) -- (C12)-- (C14);
				\draw [-] (C10) -- (C12);
				\draw[blue!75,thick] (C1) circle (7pt);
				\draw[blue!75,thick] (C2) circle (7pt);
				\draw[blue!75,thick] (C3) circle (7pt);
				\draw[blue!75,thick] (C4) circle (7pt);
				\draw[blue!75,thick] (C13) circle (7pt);
				\draw[blue!75,thick] (C14) circle (7pt);
				\draw[blue!75, rounded corners=0.2cm,thick] (-0.1,1.7)--++(-0.9,1.3)--++(0.9,1.3)--++(0.9,-1.3)--cycle;
				\draw[blue!75, rounded corners=0.2cm,thick] (-1.9,2.7)--++(-0.9,1.3)--++(0.9,1.3)--++(0.9,-1.3)--cycle;
				\end{tikzpicture}
			\end{center}
		\end{minipage}
		\begin{minipage}{0.45\textwidth}
			\begin{center}
				\tikzstyle{place}=[circle,draw=black!75, fill=black!75]
				\begin{tikzpicture}[inner sep=0.75mm,scale=0.9, every node/.style={scale=0.9}]				
				\node at (0,0) (C1) [place, label={below:${[C_1]_\delta}$}] {};  			
				\node at (-1.5,1) (C2) [place, label={left:${[C_2]_\delta}$}] {};
				\node at (0,1) (C3) [place, label={left:${[C_3]_\delta}$}] {};
				\node at (1.5,1) (C4) [place, label={right:${[C_4]_\delta}$}] {};
				\node at (-1.5,3) (C11) [place, label={left:${[C_{11}]_\delta}$}] {};
				\node at (0,3) (C12) [place, label={left:${[C_{12}]_\delta}$}] {};
				\node at (1.5,3) (C13) [place, label={right:${[C_{13}]_\delta}$}] {};
				\node at (0,4) (C14) [place, label={above:${[C_{14}]_\delta}$}] {};
				
				\draw [-] (C1) -- (C2)-- (C11)-- (C14);
				\draw [-] (C1) -- (C3)--(C11);
				\draw [-] (C3)--(C12);
				\draw [-] (C1) -- (C4)-- (C11);
				\draw [-] (C3) -- (C13)-- (C14);
				\draw [-] (C12)-- (C14);
				\draw [-] (C4) -- (C12);
				\draw [-] (C4) -- (C13);
				\end{tikzpicture}
			\end{center}
		\end{minipage}
		\caption{The local congruence $\delta$ on $\mathcal{C}(A, B, R)$  of Example~\ref{Example2} (left) and its quotient set $(\mathcal{C}(A, B, R)/\delta, \leq_\delta)$ (right).}
		\label{Example2:fig1}
	\end{figure}
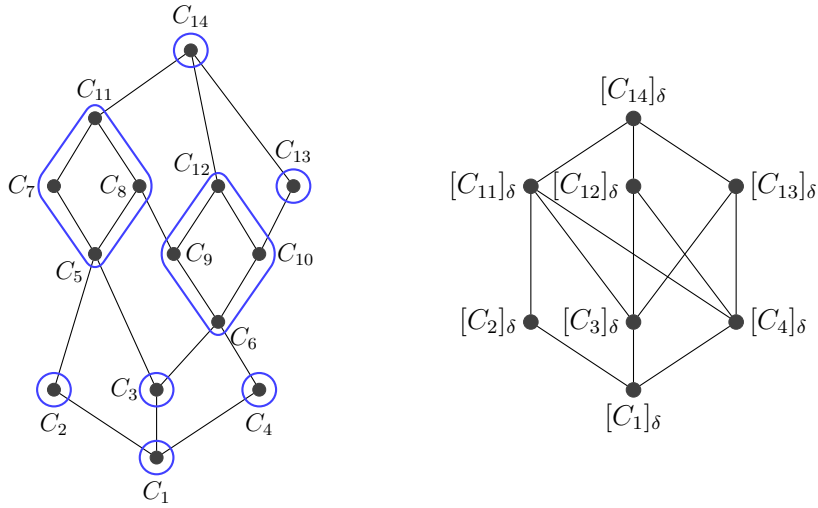
	
\end{example}

Therefore, $(\mathcal{C}(D, B, R)/\delta, \leq_\delta)$ is not a lattice in general. Moreover, local congruences can merge different equivalent classes, {which can have a remarkable impact in the reduced context with a repercussion in the reduced concept lattice.}  

Specifically,  since  $\delta$ is a  local congruence whose equivalence classes contain the classes of $(\mathcal{C}(A, B, R)/\rho_D,\sqsubseteq_D)$, given $[C]_\delta\in \mathcal{C}(A, B, R)/\delta$, there exists an index set $\Lambda{_C}$, such that 
$$
[C]_\delta=\bigcup\{[C_\lambda]_D\mid [C_\lambda]_D\in \mathcal{C}(A, B, R)/\rho_D, \lambda \in\Lambda{_C} \}
$$
where $\{[C_\lambda]_D\mid [C_\lambda]_D\in \mathcal{C}(A, B, R)/\rho_D, \lambda \in\Lambda{_C} \}$ is a convex sublattice of $(\mathcal{C}(A, B, R)/\rho_D, \sqsubseteq_D)$, in which  $[C_M]_D$ is the greatest class and $[C_m]_D$, with $C_m=\bigwedge_{C_i\in[C]_D}C_i$ is the least one.  {Hence, it is possible that some class $ [C_\lambda]_D\in \mathcal{C}(A, B, R)/\rho_D$ with $\lambda \in\Lambda{_C}$,   be a join-irreducible element. {Hence, the local congruence is} grouping     a   join-irreducible {class} into  another class. {This fact can be reflected} into the reduced context avoiding that    the concept   $ [C_\lambda]_D$   appears in the reduced concept lattice. Since  $ [C_\lambda]_D$ is join-irreducible, $\Obg(  [C_\lambda]_D)\neq \varnothing$ and this set must be removed from $B$ in order to avert the computation of this join-irreducible element. As a consequence of this deletion,} other concepts can also be removed as {a} collateral effect, as the following example shows. 

\begin{example}\label{ex:contrae0}
	We consider a context $(A, B, R)$ whose relation is given in Table~\ref{tabla:contrae0} and a subset of attribute $D =\{a_1, a_3, a_4\}$ from which we reduce the original context, that is, we remove the attributes $a_2$ and $a_5$. The associated concept lattice $\mathcal{C}(A, B, R)$  and the induced partition by the attribute reduction, $\rho_D$, are shown on Figure~\ref{contrae0:fig1}, left and right sides respectively.
	
\begin{table}[h]
\begin{center}
\begin{tabular}{l|c c c c c}
\hline
$R$ & $b_1$ & $b_2$ & $b_3$ & $b_4$ & $b_5$ \\ \hline
$a_1$ & 1 & 0 & 0 & 0 & 0 \\ 
$a_2$ & 0 & 1 & 0 & 1 & 0 \\
$a_3$ & 0 & 1 & 1 & 1 & 0 \\
$a_4$ & 1 & 0 & 0 & 1 & 0 \\
$a_5$ & 0 & 0 & 1 & 1 & 1 \\
\hline
\end{tabular}
\end{center}
\caption{Relation of Example~\ref{ex:contrae0}.
\label{tabla:contrae0}
}
\end{table}
	
	\begin{figure}[h!]
		\begin{minipage}{0.45\textwidth}
			\begin{center}
				\tikzstyle{place}=[circle,draw=black!75, fill=black!75]
				\begin{tikzpicture}[inner sep=0.75mm,scale=1, every node/.style={scale=1}]					
				\node at (1.5,2) (C1) [place, label={right:$b_1$}, label={85:$a_1$}] {}; 	
				\node at (0.5,1) (C0) [place] {}; 
				\node at (0,2) (C3) [place, label={left:$b_4$}] {};  
				\node at (-0.5,3) (C4) [place, label={left:$b_3$}] {}; 
				\node at (0.5,3) (C5) [place, label={85:$a_2$}, label={right:$b_2$}] {};
				\node at (1.5,3) (C6) [place, label={right:$a_4$}] {};
				\node at (-1.25,4) (C7) [place, label={100:$a_5$}, label={left:$b_5$}] {}; 
				\node at (0,4) (C8) [place, label={right:$a_3$}] {};   
				\node at (0,5) (C9) [place] {};	
				
				\draw [-] (C0)--(C3)--(C4)--(C8)--(C9)--(C6)--(C3)--(C5)--(C8);
				\draw [-] (C0)--(C1)--(C6);
				\draw [-] (C4) -- (C7)--(C9);
				
				\end{tikzpicture}
			\end{center}
		\end{minipage}
		\begin{minipage}{0.45\textwidth}
			\begin{center}
				\tikzstyle{place}=[circle,draw=black!75, fill=black!75]
				\begin{tikzpicture}[inner sep=0.75mm,scale=1, every node/.style={scale=1}]		
				\node at (1.5,2) (C1)  [place, label={[label distance=0.1cm]right:$C_{1}$}] {}; 	
				\node at (0.5,1) (C0) [place, label={[label distance=0.1cm]right:$C_{0}$}] {}; 
				\node at (0,2) (C2) [place, label={[label distance=0.1cm]left:$C_{2}$}] {};  
				\node at (-0.5,3) (C4) [place, label={[label distance=0.1cm]left:$C_{4}$}] {}; 
				\node at (0.5,3) (C5) [place, label={right:$C_{5}$}] {};
				\node at (1.5,3) (C3) [place, label={[label distance=0.1cm]right:$C_{3}$}] {};
				\node at (-1.25,4) (C7) [place, label={[label distance=0.1cm]left:$C_{7}$}] {}; 
				\node at (0,4) (C6) [place, label={right:$C_{6}$}] {};   
				\node at (0,5) (C8) [place, label={[label distance=0.1cm]above:$C_{8}$}] {};	
				
				\draw [-] (C0)--(C2)--(C4)--(C6)--(C8)--(C3)--(C2)--(C5)--(C6);
				\draw [-] (C0)--(C1)--(C3);
				\draw [-] (C4) -- (C7)--(C8);
				\draw[blue!75,thick, densely dashed] (C2) circle (5pt);
				\draw[blue!75,thick, densely dashed] (C3) circle (5pt);
				\draw[blue!75,thick, densely dashed] (C0) circle (5pt);
				\draw[blue!75,thick, densely dashed] (C1) circle (5pt);
				\draw[blue!75,rotate around={40:(C7)}, rounded corners=0.2cm,thick, densely dashed] (-1.4,3.8)  rectangle (0.5,4.15);
				\draw[blue!75, rounded corners=0.2cm,thick, densely dashed] (-0.7,2.85)--++(0.7,1.4)--++(0.7,-1.4)--cycle;
				\end{tikzpicture}
			\end{center}
		\end{minipage}
		\caption{The associated concept lattice $\mathcal{C}(A, B, R)$ (left) and the induced partition  $\rho_D$ (right) of Example~\ref{ex:contrae0}.}
		\label{contrae0:fig1}
	\end{figure}
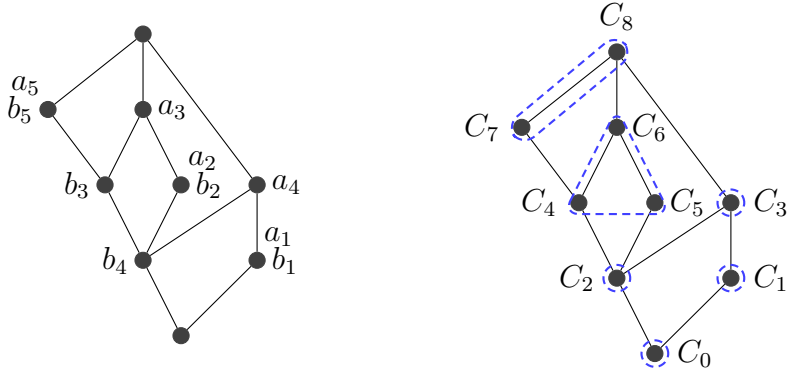
	Now, we consider the least local congruence $\delta$ on the associated concept lattice $\mathcal{C}(A, B, R)$ containing the partition induced by the attribute reduction, {which} is illustrated on the left side of Figure~\ref{contrae0:fig2}. Note that the class $[C_6]_{\delta}$ contains the class $[C_6]_D$, which is not a convex sublattice of the original concept lattice and it also contains the   {join-irreducible}  class $[C_2]_D$, as it is shown on the right side of Figure~\ref{contrae0:fig2}, due to the equivalence class of $\delta$ has to be necessarily a convex sublattice. 
	
	\begin{figure}[h!]
		\begin{minipage}{0.45\textwidth}
			\begin{center}
				\tikzstyle{place}=[circle,draw=black!75, fill=black!75]
				\begin{tikzpicture}[inner sep=0.75mm,scale=1, every node/.style={scale=1}]				
				\node at (1.5,2) (C1)  [place, label={[label distance=0.1cm]right:$C_{1}$}] {}; 	
				\node at (0.5,1) (C0) [place, label={[label distance=0.1cm]right:$C_{0}$}] {}; 
				\node at (0,2) (C2) [place, label={[label distance=0.1cm]left:$C_{2}$}] {};  
				\node at (-0.5,3) (C4) [place, label={[label distance=0.1cm]left:$C_{4}$}] {}; 
				\node at (0.5,3) (C5) [place, label={right:$C_{5}$}] {};
				\node at (1.5,3) (C3) [place, label={[label distance=0.1cm]right:$C_{3}$}] {};
				\node at (-1.25,4) (C7) [place, label={[label distance=0.1cm]left:$C_{7}$}] {}; 
				\node at (0,4) (C6) [place, label={right:$C_{6}$}] {};   
				\node at (0,5) (C8) [place, label={[label distance=0.1cm]above:$C_{8}$}] {};	
				
				\draw [-] (C0)--(C2)--(C4)--(C6)--(C8)--(C3)--(C2)--(C5)--(C6);
				\draw [-] (C0)--(C1)--(C3);
				\draw [-] (C4) -- (C7)--(C8);

				\draw[blue!75,thick] (C3) circle (5pt);
				\draw[blue!75,thick] (C0) circle (5pt);
				\draw[blue!75,thick] (C1) circle (5pt);
				\draw[blue!75, rotate around={40:(C7)}, rounded corners=0.2cm,thick] (-1.4,3.8)  rectangle (0.5,4.15);
				\draw[blue!75, rounded corners=0.2cm,thick] (-0.7,2.85)--++(0.7,1.4)--++(0.65,-1.3)--++(-0.65,-1.3)--cycle;
				\end{tikzpicture}
			\end{center}
		\end{minipage}
		\begin{minipage}{0.45\textwidth}
			\begin{center}
				\tikzstyle{place}=[circle,draw=black!75, fill=black!75]
				\begin{tikzpicture}[inner sep=0.75mm,scale=1, every node/.style={scale=1}] 	
				\node at (2,2) (C1)  [place, label={right:${[C_{1}]_D}$}] {}; 	
				\node at (1,1) (C0) [place, label={275:${[C_{0}]_D}$}] {}; 
				\node at (0,2) (C2) [place, label={[label distance=0.1cm]below:${[C_{2}]_D}$}] {};  
				\node at (1,3) (C3) [place, label={85:${[C_{3}]_D}$}] {};
				\node at (-1,3) (C6) [place, label={100:${[C_{6}]_D}$}] {};   
				\node at (0,4) (C8) [place, label={[label distance=0.1cm]above:${[C_{8}]_D}$}] {};

				\draw [-]  (C0)--(C2)--(C6)--(C8)--(C3)--(C2);
				\draw [-] (C3)--(C1)--(C0);
				\draw[blue!75,thick] (C0) circle (5pt);
				\draw[blue!75,thick] (C1) circle (5pt);
				\draw[blue!75,thick] (C3) circle (5pt);
				\draw[blue!75,thick] (C8) circle (5pt);
				\draw[blue!75, rotate around={135:(C2)}, rounded corners=0.2cm,thick] (1.6,1.8)  rectangle (-0.2,2.2);
				\end{tikzpicture}
			\end{center}
		\end{minipage}
		\caption{The local congruence $\delta$ on $\mathcal{C}(A, B, R)$  of Example~\ref{ex:contrae0} (left) and {its effect} on $\mathcal{C}(A, B, R)/\rho_D$ (right).}
		\label{contrae0:fig2}
	\end{figure}
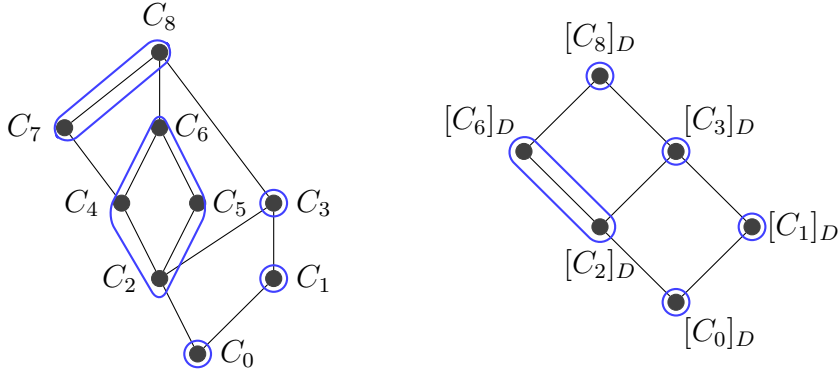

{Therefore, in order to avoid the computation of this concept,      the reduced context must be modified  removing  $\Obg(  [C_2]_D)=\{b_4\}$ from $B$. However, this modification also affects to another concept. Specifically, since $[C_3]_D = [C_1]_D \vee [C_2]_D$ and $C_3$ is not generated by any object, this concept also disappears. Thus,  grouping/removing join-irreducible elements by a local congruence  can also have some impact in other concepts. 

On the right side of Figure~\ref{contrae0:fig3},   the quotient set of the {least} local congruence $\delta$ {containing the equivalence relation associated with the reduction given by the subset $D\subseteq A$}
is depicted, and the modified context $\mathcal{C}(D,B\setminus{\{b_4\}},R_{|D\times B\setminus{\{b_4\}}})$ is shown on the left side. It is clear that both lattices are not isomorphic:
$$
\mathcal{C}(D,B,R_{|D\times B})\setminus{\{[C_2]_D\}}
\not \cong 
\mathcal{C}(D,B\setminus{\{b_4\}},R_{|D\times B\setminus{\{b_4\}}})
$$
\qed}

	 	\begin{figure}[h!]
	 	\begin{minipage}{0.45\textwidth}
	 		\begin{center}
	 			\tikzstyle{place}=[circle,draw=black!75, fill=black!75]
	 			\begin{tikzpicture}[inner sep=0.75mm,scale=1, every node/.style={scale=1}]				
	 			\node at (2,2) (C1)  [place, label={right:${C_{1}'}$}] {}; 	
				\node at (1,1) (C0) [place, label={275:${C_{0}'}$}] {};  			
				\node at (0,2) (C6) [place, label={left:${C_{6}'}$}] {};   
				\node at (1,3) (C8) [place, label={[label distance=0.1cm]above:${C_{8}'}$}] {}; 	
				
				\draw [-]  (C0)--(C6)--(C8)--(C1);
				\draw [-] (C1)--(C0);
	 
	 			\end{tikzpicture}
	 		\end{center}
	 	\end{minipage}
	 	\begin{minipage}{0.45\textwidth}
	 		\begin{center}
	 			\tikzstyle{place}=[circle,draw=black!75, fill=black!75]
	 			\begin{tikzpicture}[inner sep=0.75mm,scale=1, every node/.style={scale=1}] 	
	 			\node at (2,2) (C1)  [place, label={right:${[C_{1}]_\delta}$}] {}; 	
	 			\node at (1,1) (C0) [place, label={275:${[C_{0}]_\delta}$}] {};  
	 			\node at (2,3) (C3) [place, label={85:${[C_{3}]_\delta}$}] {};
	 			\node at (0,2.5) (C6) [place, label={left:${[C_{6}]_\delta}$}] {};   
	 			\node at (1,4) (C8) [place, label={[label distance=0.1cm]above:${[C_{8}]_\delta}$}] {};

	 			\draw [-]  (C0)--(C6)--(C8)--(C3);
	 			\draw [-] (C3)--(C1)--(C0);
	 			\end{tikzpicture}
	 		\end{center}
	 	\end{minipage}
	 	\caption{The concept lattice $(\mathcal{C}(D,B\setminus{\{b_4\}},R_{|D\times B\setminus{\{b_4\}}}),\leq_D)$ (left) and the quotient set $(\mathcal{C}(D, B, R)/\delta, \leq_{\delta})$ (right) of Example~\ref{ex:contrae0}.}
	 	\label{contrae0:fig3}
	 \end{figure}
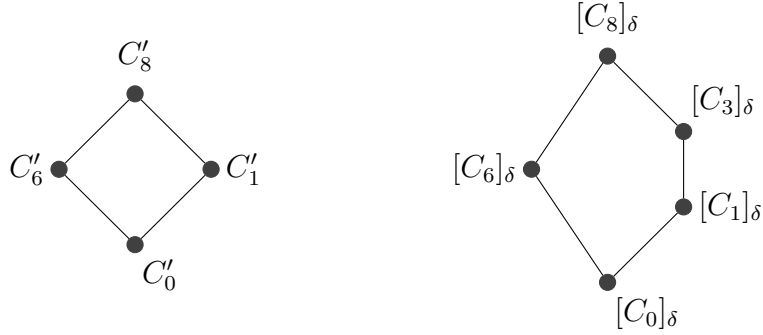
\end{example}

As we have  shown {previously}, the grouping of concepts by a local congruence can be seen as an elimination of concepts of the concept lattice, which has an impact on the context. The following section focuses on this issue.  

\section{Impact of removing elements in a concept lattices}\label{imrem:elemconcep}

{In this section, we focus on the impact of local congruences that complement attribute reductions in formal contexts. Specifically, {on the least local congruence containing  the   equivalence relation (partition) induced by   the reduction of a context $(A,B,R)$ by a subset $D\subseteq A$.
 Hence,} we are interested in analyzing how the application of this local congruence disrupts the concept lattice associated with the reduced context {$(D,B,R_{|D\times B})$}.
 
Therefore}, this section will study the necessary modifications to be done in a context, if some elements of the concept lattice are removed. For that purpose, we will distinguish two different situations described in the following two sections, {depending on whether} the removed element is a join-irreducible element or not.

{Due to local congruences group equivalence classes of the complete lattice $(\mathcal{C}(A, B, R)/\rho_D,\sqsubseteq_D)$ and, by Theorem~\ref{th:isomorph}, they can be seen as elements in $(\mathcal{C}(D, B, R_{|D\times B}),\leq_D)$, this section will be focused on the reduced context. Moreover, in order to simplify the notation, we will simply write  $(A,B,R)$ instead of the reduced context.}

\subsection{Removing join-irreducible elements}

First of all, we will study the necessary modifications in a context when  join-irreducible {elements} need to be removed preserving the rest of concepts, included the ones ``generated'' by the removed join-irreducible element, e.g. preserving the concept whose join-decomposition contains the  removed join-irreducible concept.

From now on,  we will assume that the   lattice $(\mathcal{C}(A, B, R), \leq)$ satisfies the ACC.  
As we commented above,   the subtraction of  a join-irreducible {concept $C$,  removing the objects in $\Obg(C)$,   
can also delete other concepts depending on it. However, the only subtraction of  the join-irreducible element does not alter} the structure of complete lattice. 
\begin{lemma}\label{lem:joinless}
 Given a complete lattice $(L,\wedge,\vee)$ satisfying the ACC, and $p\in L$ a join-irreducible element. Then  $(L\setminus\{p\},\wedge_p,\vee_p)$ is a sublattice of   $(L,\wedge,\vee)$, where $\wedge_p,\vee_p$ are the restriction of $\wedge,\vee$ to $L\setminus\{p\}$, and in particular a complete lattice. 
\end{lemma}
\begin{proof}
 Given $X\subseteq L\setminus\{p\}$, in particular, we have that $X\subseteq L$. Therefore, $\bigvee X$ exists in $L$. If $\bigvee X\neq p$, then the supremum also exits in $ L\setminus\{p\}$. Otherwise, if  $\bigvee X= p$, by hypothesis and Theorem~\ref{th:2.41DP}, there exists a finite subset $F\subseteq X$ such that $p= \bigvee X= \bigvee F$, which contradicts that $p$ is a join-irreducible element due to $p\not\in F$. 
Therefore,  $(L\setminus\{p\},\wedge_p,\vee_p)$ is a join-structure. Since the bottom of the lattice  $(L,\wedge,\vee)$ is not a join-irreducible  element, then it also belongs to $L\setminus\{p\}$. Thus, $(L\setminus\{p\},\wedge_p,\vee_p)$ is a sublattice of   $(L,\wedge,\vee)$.
 \qed
\end{proof}
 
Although the structure is preserved in a general lattice, {in a concept lattice, }removing a  join-irreducible element  implies  the elimination of  some objects {of the context}, which can have some impact in other concepts, as Example~\ref{ex:contrae0} shown. 
Now, we introduce a procedure to modify the original context in order to obtain a new concept lattice isomorphic to the complete lattice obtained after removing a join-irreducible {element}. 

 Given a    
 concept $ C_k\in \mathcal{C}(A, B, R)$, different from the top element, the set 
 $$
  \mathcal T(C_k)=\{C_i \mid  C_i\in   \mathcal{C}(A, B, R),   C_k< C_i\}
 $$
 is not empty, since it contains the top element $(B,B^\uparrow)$, and there exists {its} infimum, because  $(\mathcal{C}(A, B, R), \leq)$ is   a complete lattice. This infimum concept will be denoted as $C_k^t$ and the minimal elements  of $  \mathcal T(C_k)\setminus \{C_k^t\}$ as $C_k^{m_i}$, with $i$ in an index set $\Gamma$.

Now, we analyze the associated context to the concept lattice obtained after removing a join-irreducible  
 concept $ C_{j}\in \mathcal{C}(A, B, R)$. {In addition, we will denote the set of all join-decompositions of a concept $C\in\mathcal{C}(A,B,R)$ by $\mathfrak{J}(C)$.}

\begin{enumerate}
 \item\label{s1} If  $C_j^t\neq C_j$ and  
\begin{enumerate}
\item  there exists a join-decomposition of $C_j^t$, which does not contain to $C_j$, then the elements $\Obg(C_j)$ are removed from   $B$, obtaining a new set of  objects $B^*=B\setminus \Obg(C_j)$.
\item all join-decompositions of  $C_j^t$ contains to $C_j$. In this case, we consider again two cases:

\begin{enumerate}
\item if  $\Obg(C_j^t)\neq \varnothing$, 
 then the elements $\Obg(C_j)$ are removed from   $B$, obtaining a new object $B^*=B\setminus \Obg(C_j)$.

\item if  $\Obg(C_j^t)= \varnothing$, we change the elements $\Obg(C_j)$ in $B$ by a new one $b^*$, that is,  {we remove the elements $\Obg(C_j)$ from $B$, we define a new set of objects $B'= B\setminus \Obg(C_j)$ and we add the new $b^*$ to $B'$,}  $B^*=\{b^*\}\cup (B\setminus \Obg(C_j))$. Moreover, a new relation $R^*\subseteq  A\times B^*$ 
is considered, which is defined as follows:

$$
R^*=R_{|A\times B'}\cup\{(a,b^*)\mid a\in \mathcal I(C_j^t)\} 
$$
\end{enumerate}
\end{enumerate}
\item if  $C_j^t= C_j$, 
we consider the minimal elements $C_j^{m_i}$, with $i\in \Gamma$.  Clearly,  $C_j^{m_i}\neq C_j$, for all  $i\in \Gamma$, and we apply Step~\ref{s1} to all these minimal concepts. Specifically, we consider the subset $\Gamma'\subseteq \Gamma$, defined as 
$$
\Gamma'=\{i\in \Gamma\mid \hbox{all join-decompositions of  $C_j^{m_i}$ contains to $C_j$ and } \Obg(C_j^{m_i})=\varnothing\}
$$ 

\begin{enumerate}
 \item\label{Step2.i} if $\Gamma'=\varnothing$, then the elements $\Obg(C_j)$ are removed from   $B$, obtaining a new set of objects $B^*=B\setminus \Obg(C_j)$.
 
  \item otherwise, $\Gamma'\neq\varnothing$, and we define the new set of objects as  $B^*=\{b_i^*\mid i\in \Gamma'\}\cup {B'}$, {where $B' = B\setminus \Obg(C_j)$}. Moreover, a new relation $R^*\subseteq  A\times B^*$ 
is considered, which is defined as follows:

$$
R^*=R_{|A\times B'}\cup\{(a,b_i^*)\mid a\in \mathcal I(C_j^{{m_i}}),  i\in \Gamma'\} 
$$

\end{enumerate}

\end{enumerate}

These steps are translated into Algorithm~\ref{al:joincontext} in order to present a more simple an operational procedure. 

\begin{algorithm}[!h]
	
	\caption{Removing a join-irreducible concept {from} a concept lattice}
	\label{al:joincontext}
	
	\SetKwInOut{Input}{input}\SetKwInOut{Output}{output}
	
	\Input{$(A,B,R)$, $C_j$, $\mathcal T(C_j)$, $\Gamma$}
	\Output{$(A, B^*, R^*)$}
	\BlankLine
	
	Compute the infimum concept $C_j^t$ of the set $\mathcal T(C_j)$\;
	\If{$C_j^t \neq C_j$,}{
		\If{there exists $\chi\in \mathfrak{J}(C_j^t)$ such that $C_j\not\in \chi$ {\bf or} $\Obg(C_j^t) \neq \varnothing$ \label{Case1}}{
			$B^*=B\setminus \Obg(C_j)$\;
			$R^* = R_{|A\times B^*}$\;
		}
		\Else{\label{Case2}$B' = B\setminus \Obg(C_j)$\;
			$B^*=\{b^*\}\cup B'$\;
			$R^*=R_{|A\times B'}\cup\{(a,b^*)\mid a\in \mathcal I(C_j^t)\}$\;
			
		}
	}
	\Else{Compute the minimal elements $C_j^{m_i}$ with $i\in \Gamma$\;
		Define the subset $\Gamma'\subseteq \Gamma$ as\\
		$\Gamma'=\{i\in \Gamma\mid C_j\in \chi \hbox{ for all } \chi\in\mathfrak{J}(C_j^{m_i})\hbox{ and } \Obg(C_j^{m_i})=\varnothing\}$\;
		\If{$\Gamma' = \varnothing$\label{Case3}}{$B^*=B\setminus \Obg(C_j)$\;
			$R^* = R_{|A\times B^*}$\;
		}
		\Else{\label{Case4}$B' = B\setminus \Obg(C_j)$\;
			$B^*=\{b_i^*\mid i\in \Gamma'\}\cup B'$\;
			$R^*=R_{|A\times B'}\cup\{(a,b_i^*)\mid a\in \mathcal I(C_j^t),~ i\in \Gamma'\} $\;
		}
	}
	\BlankLine
	\Return{$(A, B^*, R^*)$}
\end{algorithm}

Notice that the first step and Step~\ref{Step2.i} are merged in Line~\ref{Case1}. 
Since $C_{j}$ is join-irreducible, if $\Obg(C_j)$ has only one element, it is an absolutely necessary object\footnote{We are considering dual notions {of} attribute reduction~\cite{Medina2016}.} and if $\Obg(C_j)$ has more than one element, they are relatively necessary objects. Hence, all of them need to be removed in order to ensure that the concept $C_j$ does not appear. 

The mechanism above computes the required context, as the following result proves. 

\begin{theorem}\label{th:removejoin}
 Given a context $(A,B,R)$ and a join-irreducible concept $C_j\in \mathcal{C}(A, B, R)$, we have that 
$$
 (\mathcal{C}(A, B, R)\setminus \{C_j\},\leq_j) \cong (\mathcal{C}(A, B^*, R^*),\leq^*), 
$$
where $B^*$ and $R^*$ {is the set and the relation} given by Algorithm~\ref{al:joincontext}, and $\leq_j$ is the ordering defined from the restriction of $\leq$ to $\mathcal{C}(A, B, R)\setminus \{C_j\}$.
 
\end{theorem}

\begin{proof}

{The proof straightforwardly holds, if the intents of both set of concepts coincide.}
First of all, we will prove that the set of intents of concepts of $\mathcal{C}(A, B^*, R^*)$ is contained in the sets of intents of concepts of $\mathcal{C}(A, B, R)\setminus \{C_j\}$. For that purpose, let us consider an arbitrary concept $(X, Y) \in \mathcal{C}(A, B^*, R^*)$ and we will prove that $Y$ is also the intent of a concept of $\mathcal{C}(A, B, R)\setminus \{C_j\}$. We differentiate the derivation operators of each context, that is, for the context $(A, B^*, R^*)$ we will denote the derivation operators as $({\ }^{\uparrow_{*}}, {\ }^{\downarrow^{*}})$ and for $(A, B, R)$ as $({\ }^\uparrow, {\ }^\downarrow)$.

According to Algorithm~\ref{al:joincontext},  several cases should be distinguished:

\begin{enumerate}
	\item If in the construction of  $(A, B^*, R^*)$ the condition given in Line~\ref{Case1} 
	has been satisfied, then $B^*\subseteq B$ and, since the set of attributes is the same in both contexts, we have that $X^{\uparrow_{ *}} = X^\uparrow$. Therefore, the following chain of inequalities holds:
	$$Y^{\downarrow\uparrow} = X^{\uparrow_{*}\downarrow\uparrow}= X^{\uparrow\downarrow\uparrow} = X^\uparrow = X^{\uparrow_{*}} = Y$$
	\item If in the construction of  $(A, B^*, R^*)$ the condition given in Line~\ref{Case2}  
	has been satisfied, then $B^* = B' \cup \{b^*\}$ where $B'\subseteq B$. In this case, we have to differentiate three cases:
	\renewcommand{\labelenumii}{\labelenumi\arabic{enumii}.} 
	\begin{enumerate}
		\item If $Y{\subseteq} b^{*\uparrow_{ *}}$, then we {have that  $b^*\in b^{*\uparrow_{ *}\downarrow_{*}} \subseteq Y^{\downarrow_{*}}=X $ and so,} the extent of the concept {can be written} as $X = X_0 \cup \{b^*\}$ where $X_0 \subseteq B'$ and, since the set of attributes is the same in both context, we have that $X_0^{\uparrow_{ *}} = X_0^\uparrow$. Thus, $ Y = X^{\uparrow_{ *}} = 
		(X_0 \cup \{b^*\})^{\uparrow_{ *}}=
		X_0^{\uparrow_{ *}} \cap b^{*\uparrow_{ *}}= X_0^{\uparrow} \cap \inte(C_j^t)$ by construction in Algorithm~\ref{al:joincontext} and therefore,
			$$Y^{\downarrow\uparrow} = X^{\uparrow_{*}\downarrow\uparrow}= (X_0^{\uparrow} \cap \inte(C_j^t))^{\downarrow\uparrow} \overset{(*)}{=} (X_0 \cup \ext(C_j^t))^{\uparrow\downarrow\uparrow} = (X_0 \cup \ext(C_j^t))^{\uparrow} $$
			The equality $(*)$ holds because $C_j^t$ is a concept { of $(A, B, R)$} and therefore,
			$$ (X_0 \cup \ext(C_j^t))^{\uparrow} = X_0^{\uparrow} \cap \inte(C_j^t) =  X_0^{\uparrow_{ *}} \cap b^{*\uparrow_{ *}} = X^{\uparrow_{ *}}= Y$$
		
		\item If  $Y\not\subseteq b^{*\uparrow_{ *}}$, then $X\subseteq B'\subseteq B$. Thus, the demonstration is analogous to the one given in the first case.

	\end{enumerate}
	\item If in the construction of  $(A, B^*, R^*)$ the condition given in Line~\ref{Case3} 
	has been satisfied, the proof follows an analogous reasoning to the one given in the first case.
	\item If in the construction of  $(A, B^*, R^*)$ the condition given in Line~\ref{Case4} 
	has been satisfied, the proof is analogous to one given in the second case, but considering $\{b_i^*\}$ instead of $\{b^*\}$.
\end{enumerate}

Now, we will prove that the intents of concepts of $\mathcal{C}(A, B, R)\setminus\{C_j\}$ are also intents of concepts of $ \mathcal{C}(A, B^*, R^*)$.  For that purpose, let us consider a concept $(X, Y) \in \mathcal{C}(A, B, R)\setminus\{C_j\}$. Depending on the removed concept $C_j$, we can distinguish several cases considered in Algorithm~\ref{al:joincontext}:
\begin{enumerate}
	\item  If the condition given in Line~\ref{Case1} % in Algorithm~\ref{al:joincontext} 
	is satisfied, then $B^*\subseteq B$ and, since the set of attributes is the same in both context, we have that $X^{\uparrow_{ *}} = X^\uparrow$. Therefore,
	$$Y^{\downarrow^*\uparrow_*} = X^{\uparrow\downarrow^*\uparrow_*}= X^{\uparrow_*\downarrow^*\uparrow_*} = X^{\uparrow_*} = X^{\uparrow} = Y$$
	\item If the condition given in Line~\ref{Case2} %in Algorithm~\ref{al:joincontext} 
	is satisfied, then $B^* = B' \cup \{b^*\}$ where $B'\subseteq B$. In this case, we have to differentiate three cases:
	\renewcommand{\labelenumii}{\labelenumi\arabic{enumii}.} 
	\begin{enumerate}
		\item If $Y{\subseteq} \inte(C_j^t)$, then we can see the extent as $X = X_0 \cup \ext(C_j^t)$. Thus, $ Y = X^{\uparrow} = X_0^{\uparrow} \cap \ext(C^t_j)^{\uparrow}= X_0^{\uparrow} \cap \inte(C^t_j)= X_0^{\uparrow_{*}} \cap \inte(C^t_j)$ since $X_0\subseteq B' \subseteq B^*$ and the set of attributes is the same in both context, we have that $X_0^\uparrow = X_0^{\uparrow_{ *}}$. Therefore,
			$$Y^{\downarrow^*\uparrow_*} = X^{\uparrow\downarrow^*\uparrow_*}= 
			(X_0^{\uparrow_{*}} \cap \inte(C^t_j))^{\downarrow^*\uparrow_*} \overset{(*)}{=} (X_0^{\uparrow_{*}} \cap b^{*\uparrow_{ *}})^{\downarrow^*\uparrow_*} = (X_0 \cup b^{*})^{\uparrow_*\downarrow^*\uparrow_*}$$
			The equality $(*)$ holds because $\inte(C_j^t) = b^{*\uparrow_{ *}}$ by construction in Algorithm~\ref{al:joincontext}. Therefore,
			$$(X_0 \cup b^{*})^{\uparrow_*\downarrow^*\uparrow_*} = (X_0 \cup b^{*})^{\uparrow_*}= X_0^{\uparrow_{*}} \cap b^{*\uparrow_{ *}} = X_0^{\uparrow} \cap \inte(C_j^t) =  X^{\uparrow}= Y$$
		
		\item If  $Y\not\subseteq \inte(C_j^t)$, then $X\subseteq B'\subseteq B$. Thus, the proof of this case is similar to the first case.

	\end{enumerate}
	\item If the condition given in Line~\ref{Case3} 
	is satisfied. The proof of this case is analogous to the one given in the first case.
	\item If the condition given in Line~\ref{Case4} 
	is satisfied. The proof of this case is analogous to the one given in the second case, but considering $\{b_i^*\}$ instead of $\{b^*\}$.
\end{enumerate}

Therefore, we have proved that the intents are preserved {and so, both lattices are isomorphic}.\qed
\end{proof}

This procedure can be applied sequentially, when more than one join-irreducible element need to be removed.

Moreover, by the notions of attribute and object reduction, 
the number of modified objects is the minimum one for ensuring the isomorphism in Theorem~\ref{th:removejoin}. {Notice that all objects in $\Obg(C_j)$ have been removed in order to erase $C_j$ and the minimum number of objects (only one per concept) have been introduced to preserve the concepts depending (in a join-decomposition) on $C_j$ (see Example~\ref{ex:contrae0})}. {Thus, the proposed mechanism, to characterize the impact of removing a join-irreducible concept from the concept lattice in the context, provides the closest  context to the original one.}

\subsection{Removing non-join-irreducible elements}

Moreover, we also need to inspect the possibility of removing a non-join-irreducible element, since the class $[C]_\delta$ can also includes this kind of elements. 
This section will be devoted to this issue.

In a general lattice $(L,\wedge,\vee)$, given a non-join-irreducible element  $y\in L$, it can be meet-irreducible or not. Clearly, in the former case  a dual result to Lemma~\ref{lem:joinless} arises.

\begin{lemma}\label{lem:meetless}
 Given a complete lattice $(L,\wedge,\vee)$ satisfying the DCC, and $q\in L$ a meet-irreducible element. Then  $(L\setminus\{q\},\wedge_q,\vee_q)$ is a sublattice of   $(L,\wedge,\vee)$, where $\wedge_q,\vee_q$ are the restriction of $\wedge,\vee$ to $L\setminus\{q\}$, and in particular a complete lattice. 
\end{lemma}
\begin{proof}
The proof is dual to the one given in Lemma~\ref{lem:joinless}.
 \qed
\end{proof}

Therefore, if a meet-irreducible concept is removed, a dual procedure to Algorithm~\ref{al:joincontext} can be done removing attributes instead of objects, {which is detailed in Algorithm~\ref{al:meetcontext}.} Notice that the set $\mathfrak{M}(C_k)$ is the set of all  meet-decomposition{s} of a concept $C_k\in\mathcal C(A,B,R)$ and the set $\mathcal S(C_k)$ is the dual of {the} set $\mathcal{T}(C_k)$, that is, $\mathcal S(C_k)=\{C_i \mid  C_i\in   \mathcal{C}(A, B, R),   C_i< C_k\}$. Also a dual result to {Theorem~}\ref{th:removejoin} arises when a meet-irreducible element is removed from a general context.

{\begin{theorem}\label{th:removemeet}
 Given a context $(A,B,R)$ and a meet-irreducible concept $C_k\in \mathcal{C}(A, B, R)$, we have that 
$$
 (\mathcal{C}(A, B, R)\setminus \{C_k\},\leq_k) \cong (\mathcal{C}(A^*, B, R^*),\leq^*), 
$$
where $A^*$ and $R^*$ is the set and the relation given by Algorithm~\ref{al:meetcontext}, and $\leq_k$ is the ordering defined from the restriction of $\leq$ to $\mathcal{C}(A, B, R)\setminus \{C_k\}$.
 
\end{theorem}

\begin{proof}
The proof is dual to the one given to Theorem~\ref{th:removejoin}. \qed
\end{proof}}

\begin{algorithm}[!h]
	
	\caption{Removing a meet-irreducible concept of a concept lattice}
	\label{al:meetcontext}
	
	\SetKwInOut{Input}{input}\SetKwInOut{Output}{output}
	
	\Input{$(A,B,R)$, $C_l$, $\mathcal S(C_l)$, $\Gamma$}
	\Output{$(A^*, B, R^*)$}
	\BlankLine
	Compute the supremum concept $C_l^s$ of the set $\mathcal S(C_l)$\;
	\If{$C_l^s \neq C_l$,}{\label{Step_1}
		\If{there exists $\psi\in \mathfrak{M}(C_l^s)$ such that $C_l\not\in \psi$ {\bf or} $\Atg(C_l^s) \neq \varnothing$ }{
			$A^*=A\setminus \Atg(C_l)$\;
			$R^* = R_{|A^*\times B}$\;
		}
		\Else{$A' = A\setminus \Atg(C_l)$\;
			$A^*=\{a^*\}\cup A'$\;
			$R^*=R_{|A'\times B}\cup\{(a^*,b)\mid b\in \ext(C_l^s)\}$\;
			
		}
	}
	\Else{Compute the maximal elements $C_l^{s_i}$ with $i\in \Gamma$\;
		Define the subset $\Gamma'\subseteq \Gamma$ as\\
		$\Gamma'=\{i\in \Gamma\mid C_l\in \psi \hbox{ for all } \psi\in\mathfrak{M}(C_l^{s_i})\hbox{ and } \Atg(C_l^{s_i})=\varnothing\}$\;
		\If{$\Gamma' = \varnothing$}{$A^*=A\setminus \Atg(C_l)$\;
			$R^* = R_{|A^*\times B}$\;
		}
		\Else{$A' = A\setminus \Atg(C_l)$\;
			$A^*=\{a_i^*\mid i\in \Gamma'\}\cup A'$\;
			$R^*=R_{|A'\times B}\cup\{(a_i^*,b)\mid b\in \ext(C_l^s),~ i\in \Gamma'\} $\;
		}
	}
	\BlankLine
	\Return{$(A^*, B, R^*)$}
\end{algorithm}

{Now, we will consider the case when the element to be removed 
 is neither  meet-irreducible  nor join-irreducible. The following example shows that the structure of complete lattice does not hold.} 

\begin{example}\label{ex:contrae1}
		We consider a context $(A, B, R)$ whose relation is given in Table~\ref{tabla:contrae1} and a subset of attribute{s} $D =\{a_1, a_2, a_3, a_4\}$, that is, we remove the attributes $a_5$ and $a_6$. The  concept lattice $\mathcal{C}(A, B, R)$  and the induced partition by the attribute reduction, $\rho_D$, are shown in Figure~\ref{contrae1:fig1}, left and right sides respectively.
		
		\begin{table}[h]
			\begin{center}
				\begin{tabular}{l|c c c c c c}
					\hline
					$R$ & $b_1$  & $b_2$ & $b_3$ & $b_4$ & $b_5$ & $b_6$\\ \hline
					$a_1$ & 1 & 0 & 0 & 0  & 0 & 0\\ 
					$a_2$ & 0 & 1 & 0 & 0  & 0 & 0\\
					$a_3$ & 1 & 1 & 1 & 0  & 1 & 0\\
					$a_4$ & 1 & 1 & 0 & 1  & 0 & 0\\
					$a_5$ & 1 & 1 & 0 & 0  & 1 & 0\\
					$a_6$ & 1 & 1 & 1 & 0  & 0 & 1\\
					\hline
				\end{tabular}
			\end{center}
			\caption{Relation of Example~\ref{ex:contrae1}.
				\label{tabla:contrae1}
			}
		\end{table}
		
		\begin{figure}[h!]
			\begin{minipage}{0.45\textwidth}
				\begin{center}
					\tikzstyle{place}=[circle,draw=black!75, fill=black!75]
					\begin{tikzpicture}[inner sep=0.75mm,scale=1, every node/.style={scale=1}]				
					\node at (0,0) (C0) [place] {};  	
					\node at (-0.5,1) (C1) [place, label={left:$b_2$}, label={100:$a_2$}] {}; 	
					\node at (0.5,1) (C2) [place, label={right:$b_1$}, label={85:$a_1$}] {}; 
					\node at (0,2) (C3) [place] {};  
					\node at (-0.5,3) (C4) [place, label={left:$b_3$}] {}; 
					\node at (0.5,3) (C5) [place, label={85:$a_5$}, label={right:$b_5$}] {};
					\node at (1.5,3) (C6) [place, label={85:$a_4$}, label={right:$b_4$}] {};
					\node at (-1.25,4) (C7) [place, label={100:$a_6$}, label={left:$b_6$}] {}; 
					\node at (0,4) (C8) [place, label={right:$a_3$}] {};   
					\node at (0,5) (C9) [place] {};

					\draw [-] (C0)--(C1)--(C3)--(C4)--(C8)--(C9)--(C6)--(C3)--(C5)--(C8);
					\draw [-] (C0) -- (C2)--(C3);
					\draw [-] (C4) -- (C7)--(C9);
					
					\end{tikzpicture}
				\end{center}
			\end{minipage}
			\begin{minipage}{0.45\textwidth}
				\begin{center}
					\tikzstyle{place}=[circle,draw=black!75, fill=black!75]
					\begin{tikzpicture}[inner sep=0.75mm,scale=1, every node/.style={scale=1}]	
					\node at (0,0) (C0) [place, label={[label distance=0.1cm]left:$C_{0}$}] {};  	
					\node at (-0.5,1) (C1)  [place, label={[label distance=0.1cm]left:$C_{1}$}] {}; 	
					\node at (0.5,1) (C2) [place, label={[label distance=0.1cm]right:$C_{2}$}] {}; 
					\node at (0,2) (C3) [place, label={[label distance=0.1cm]left:$C_{3}$}] {};  
					\node at (-0.5,3) (C4) [place, label={[label distance=0.1cm]left:$C_{4}$}] {}; 
					\node at (0.5,3) (C5) [place, label={[label distance=0.1cm]right:$C_{5}$}] {};
					\node at (1.5,3) (C6) [place, label={[label distance=0.1cm]right:$C_{6}$}] {};
					\node at (-1.25,4) (C7) [place, label={[label distance=0.1cm]left:$C_{7}$}] {}; 
					\node at (0,4) (C8) [place, label={[label distance=0.1cm]right:$C_{8}$}] {};   
					\node at (0,5) (C9) [place, label={[label distance=0.1cm]above:$C_{9}$}] {};	
					
					\draw [-] (C0)--(C1)--(C3)--(C4)--(C8)--(C9)--(C6)--(C3)--(C5)--(C8);
					\draw [-] (C0) -- (C2)--(C3);
					\draw [-] (C4) -- (C7)--(C9);
					\draw[blue!75,thick, densely dashed] (C3) circle (5pt);
					\draw[blue!75,thick, densely dashed] (C2) circle (5pt);
					\draw[blue!75,thick, densely dashed] (C6) circle (5pt);
					\draw[blue!75,thick, densely dashed] (C0) circle (5pt);
					\draw[blue!75,thick, densely dashed] (C1) circle (5pt);
					\draw[blue!75, rotate around={40:(C7)}, rounded corners=0.2cm,thick, densely dashed] (-1.4,3.8)  rectangle (0.5,4.15);
					\draw[blue!75, rounded corners=0.2cm,thick, densely dashed] (-0.7,2.85)--++(0.7,1.4)--++(0.7,-1.4)--cycle;
					\end{tikzpicture}
				\end{center}
			\end{minipage}
			\caption{The   concept lattice $\mathcal{C}(A, B, R)$ (left) and the induced partition  $\rho_D$ (right) of Example~\ref{ex:contrae1}.}
			\label{contrae1:fig1}
		\end{figure}
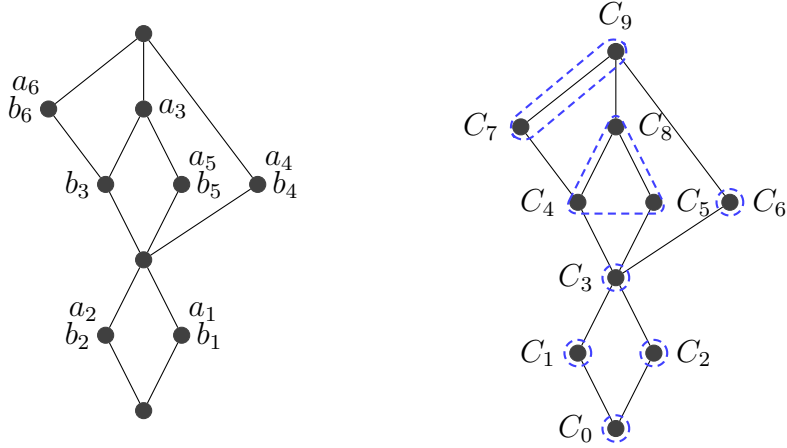
		Now, we consider the least local congruence $\delta$ on the   concept lattice $\mathcal{C}(A, B, R)$ containing the partition induced by the attribute reduction, {which} is illustrated in the left side of Figure~\ref{contrae1:fig2}. Note that the class $[C_8]_{\delta}$ contains the class $[C_8]_D$, which is not a convex sublattice of the original concept lattice and {so,} it also contains the class $[C_3]_D$, as it is shown in the middle of Figure~\ref{contrae1:fig2}, due to the equivalence class of $\delta$ has to be necessarily a convex sublattice. 
		
		\begin{figure}[h!]
			\begin{minipage}{0.32\textwidth}
				\begin{center}
					\tikzstyle{place}=[circle,draw=black!75, fill=black!75]
					\begin{tikzpicture}[inner sep=0.75mm,scale=1, every node/.style={scale=1}]				
					\node at (0,0) (C0) [place, label={[label distance=0.1cm]left:$C_{0}$}] {};  	
					\node at (-0.5,1) (C1)  [place, label={[label distance=0.1cm]left:$C_{1}$}] {}; 	
					\node at (0.5,1) (C2) [place, label={[label distance=0.1cm]right:$C_{2}$}] {}; 
					\node at (0,2) (C3) [place, label={[label distance=0.1cm]left:$C_{3}$}] {};  
					\node at (-0.5,3) (C4) [place, label={[label distance=0.1cm]left:$C_{4}$}] {}; 
					\node at (0.5,3) (C5) [place, label={[label distance=0.1cm]right:$C_{5}$}] {};
					\node at (1.5,3) (C6) [place, label={[label distance=0.1cm]right:$C_{6}$}] {};
					\node at (-1.25,4) (C7) [place, label={[label distance=0.1cm]left:$C_{7}$}] {}; 
					\node at (0,4) (C8) [place, label={[label distance=0.1cm]right:$C_{8}$}] {};   
					\node at (0,5) (C9) [place, label={[label distance=0.1cm]above:$C_{9}$}] {};	
					
					\draw [-] (C0)--(C1)--(C3)--(C4)--(C8)--(C9)--(C6)--(C3)--(C5)--(C8);
					\draw [-] (C0) -- (C2)--(C3);
					\draw [-] (C4) -- (C7)--(C9);
					\draw[blue!75,thick] (C2) circle (5pt);
					\draw[blue!75,thick] (C6) circle (5pt);
					\draw[blue!75,thick] (C0) circle (5pt);
					\draw[blue!75,thick] (C1) circle (5pt);
					\draw[blue!75,rotate around={40:(C7)}, rounded corners=0.2cm,thick] (-1.4,3.8)  rectangle (0.5,4.15);
					\draw[blue!75, rounded corners=0.2cm,thick] (-0.7,2.85)--++(0.7,1.4)--++(0.65,-1.3)--++(-0.65,-1.3)--cycle;
					\end{tikzpicture}
				\end{center}
			\end{minipage}
			\begin{minipage}{0.33\textwidth}
				\begin{center}
					\tikzstyle{place}=[circle,draw=black!75, fill=black!75]
					\begin{tikzpicture}[inner sep=0.75mm,scale=1, every node/.style={scale=1}]				
					\node at (0,0) (C0) [place, label={[label distance=0.1cm]below:${[C_{0}]_D}$}] {};  	
					\node at (-1,1) (C1)  [place, label={100:${[C_{1}]_D}$}] {}; 	
					\node at (1,1) (C2) [place, label={275:${[C_{2}]_D}$}] {}; 
					\node at (0,2) (C3) [place, label={[label distance=0.1cm]right:${[C_{3}]_D}$}] {};  
					\node at (1,3) (C6) [place, label={85:${[C_{6}]_D}$}] {};
					\node at (-1,3) (C8) [place, label={100:${[C_{8}]_D}$}] {};   
					\node at (0,4) (C9) [place, label={[label distance=0.1cm]above:${[C_{9}]_D}$}] {};

					\draw [-] (C0)--(C1)--(C3)--(C8)--(C9)--(C6)--(C3)--(C2)--(C0);
					\draw[blue!75,thick] (C2) circle (5pt);
					\draw[blue!75,thick] (C0) circle (5pt);
					\draw[blue!75,thick] (C1) circle (5pt);
					\draw[blue!75,thick] (C6) circle (5pt);
					\draw[blue!75,thick] (C9) circle (5pt);
					\draw[blue!75, rotate around={135:(C3)}, rounded corners=0.2cm,thick] (1.6,1.8)  rectangle (-0.2,2.2);
					\end{tikzpicture}
				\end{center}
			\end{minipage}
			\begin{minipage}{0.33\textwidth}
				\tikzstyle{place}=[circle,draw=black!75, fill=black!75]
				\begin{tikzpicture}[inner sep=0.75mm,scale=1, every node/.style={scale=1}]				
				\node at (0,0) (C0) [place, label={below:${[C_{0}]_\delta}$}] {};  	
				\node at (-1,1) (C1)  [place, label={265:${[C_{1}]_\delta}$}] {}; 	
				\node at (1,1) (C2) [place, label={275:${[C_{2}]_\delta}$}] {}; 
				\node at (1,3) (C6) [place, label={85:${[C_{6}]_\delta}$}] {};
				\node at (-1,3) (C8) [place, label={96:${[C_{8}]_\delta}$}] {};   
				\node at (0,4) (C9) [place, label={above:${[C_{9}]_\delta}$}] {};

				\draw [-] (C0)--(C1)--(C8)--(C9)--(C6)--(C1);
				\draw [-] (C0)--(C2)--(C8)--(C9)--(C6)--(C2);
				\end{tikzpicture}
			\end{minipage}
			\caption{The local congruence $\delta$ on $\mathcal{C}(A, B, R)$ (left), $\delta$ on $\mathcal{C}(A, B, R)/\rho_D$ (middle) and the quotient set $(\mathcal{C}(D, B, R)/\delta, \leq_\delta)$ of Example~\ref{ex:contrae1}.}
			\label{contrae1:fig2}
		\end{figure}
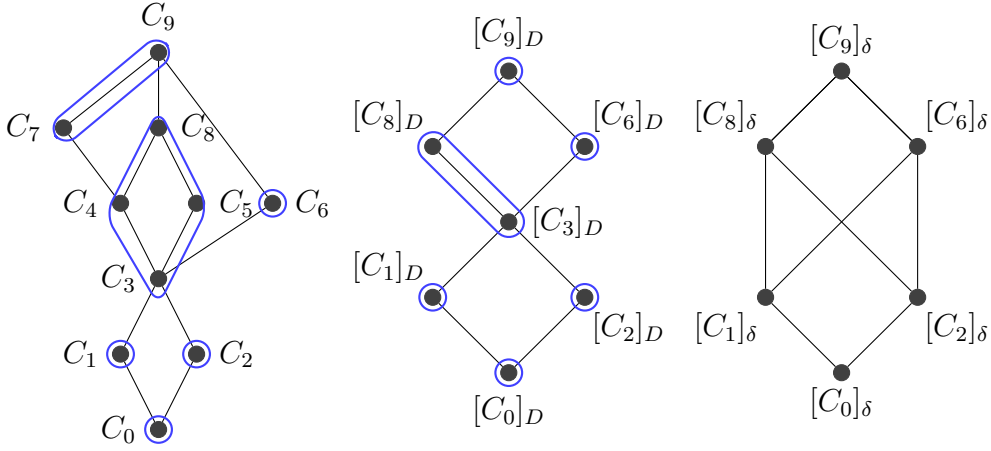
		
		However, the class $[C_3]_D$ is neither a meet-irreducible nor {a} join-irreducible element of $\mathcal{C}(A, B, R)/\rho_D$ and if we merge it with $[C_8]_D$ as the local congruence $\delta$ does, then the lattice structure is broken. As a consequence, the quotient set $(\mathcal{C}(D,B,R)/\delta, \leq_\delta)$ is not a lattice but {only} a poset, as it is shown in the right side of Figure~\ref{contrae1:fig2}.  \qed
\end{example}

Although the obtained poset is not a lattice, the Dedekind-MacNeille completion of the obtained poset provides the structure of {a} complete lattice. Indeed, this complete lattice is isomorphic to the  lattice in which we have decided to remove a concept. 

\begin{lemma}\label{lem:joinmeetless}
 Given a complete lattice 
$(L,\preceq )$ {satisfying the ACC and DCC}, and a non-meet-irreducible  and non-join-irreducible element $y\in L$. Then  
$$
(\DM(L\setminus\{y\}),\subseteq)\cong (L,\preceq)
$$
\end{lemma}
\begin{proof}
{Since $(L,\preceq )$ satisfies the ACC and DCC, then clearly $L\setminus\{y\}$ is join-dense and meet-dense (it contains the whole set of join and meet irreducible elements of $L$) and so,  by Theorem~\ref{DP:Th7.42}, we obtain the  isomorphism  $
(\DM(L\setminus\{y\}),\subseteq)\cong (L,\preceq)
$.
 \qed}
\end{proof}

Therefore, since the attribute reduction procedure based on local congruences is also focused on obtaining a quotient set {with the structure of a complete lattice,}    { the    Dedekind-MacNeille completion can be applied to achieve this challenge.}   

{\begin{proposition}
 Given a context $(A,B,R)$ and a non-join and non-meet-irreducible concept $C_i\in \mathcal{C}(A, B, R)$, we have that 
$$
 (\DM(\mathcal{C}(A, B, R)\setminus \{C_i\}),\subseteq_i) \cong (\mathcal{C}(A, B, R),\leq), 
$$
where $\subseteq_i$ is the ordering defined by the Dedekind-MacNeille completion.
 
\end{proposition}

\begin{proof}
The proof straightforwardly follows from Lemma~\ref{lem:joinmeetless}.\qed
\end{proof}}

{
Therefore, although a class with only this kind of concepts will be grouped in another class, no impact have in the concept lattice. Therefore, the application of the Dedekind-MacNeille completion is only needed to obtain a complete lattice isomorphic to the original one. The following theorem summarizes these results.}

{Similarly to the other procedures, this mechanism can be applied sequentially to different concepts and, moreover, Algorithms~\ref{al:joincontext} and~\ref{al:meetcontext} can be interspersed.} 

{\begin{theorem}\label{cor:noyo}
 Given a context $(A,B,R)$ and a concept $C_k\in \mathcal{C}(A, B, R)$, we have that 
$$
 (\DM(\mathcal{C}(A, B, R)\setminus \{C_k\}),\subseteq_k) \cong (\mathcal{C}(A^*, B^*, R^*),\leq^*), 
$$
where $A^*$, $B^*$ and $R^*$ are the sets and {the} relation given by either  Algorithm~\ref{al:joincontext} or Algorithm~\ref{al:meetcontext} or the original ones in case of $C_k$ is neither {a} join nor {a}  meet irreducible element, and $\subseteq_k$ is the ordering defined by the Dedekind-MacNeille completion.

\end{theorem}}
{\begin{proof}
 The proof follows from Theorems~\ref{th:removejoin} and~\ref{th:removemeet}, and the comment above concerning neither  join nor    meet irreducible elements.\qed

\end{proof}}
{Since this theorem can be applied sequentially to the classes grouped by a local congruence, we have just characterized the impact of complementing an attribute reduction with the application of a local congruence. As a consequence, we can obtain a context from the original one which is isomorphic to the quotient set of the local congruence, considering the ordering defined in Theorem~\ref{newpo}. 
It is also relevant  to highlight that  this mechanism computes the contexts more similar to the original one.} Notice that either Algorithm~\ref{al:joincontext} or Algorithm~\ref{al:meetcontext} can be applied to meet and join irreducible concepts and so, two different (although isomorphic) contexts arise. Therefore, the user can decide what kind of elements (attributes or objects) prefers to  modify, and obtain  the closest context to the original one, under this criterion.

\section{Conclusions and future work}\label{conclusion}

In this work, we have studied the impact of applying a local congruence on a concept lattice associated with a reduced context. We have proved that the quotient set generated by the equivalence relation induced by an attribute reduction is isomorphic to the concept lattice corresponding to the the reduced context. However, we have seen that this fact does not hold for the quotient set generated by local congruences. We have also shown that   the {clustering} carried out by a local congruence, {after an attribute reduction,}  can  have some impact in other concepts of the concept lattice. For that reason, we have studied the necessary modification to be done in a context when a concept of an arbitrary concept lattice needs to be removed. For this study, we have distinguished two types of elements in the lattice: join-irreducible and non-join-irreducible elements. We have proved that when we remove a join-irreducible element from a {general} complete lattice, the structure of complete lattice is preserved. Furthermore, it has been presented and proved a procedure for computing a modified context whose associated concept lattice is isomorphic to the original concept lattice  when one of its join-irreducible {concepts} has been removed {throughout the delation of the objects generating that concept}. In addition, dual results  can be obtained for meet-irreducible concepts and an analogous procedure has been introduced when the removed element is a meet-irreducible concept. Finally, we have also analyzed the case when the removed element is neither {a} meet nor  {a} join-irreducible element of the concept lattice, showing that, in this particular case,  the Dedekind-MacNeille completion of the obtained poset is needed in order to provide the structure of a complete lattice.

In the near future, {the introduced algorithms will be complemented with different attribute reduction mechanisms~\cite{Cornejo2017,TFS:2020-acmr,konecny2019} and will be applied to real databases, such as the ones collected from our participation in the COST Action DigForASP. Moreover, the relationship between local congruences and attribute implications will be  studied.}

% ---- Bibliography ----
%
% BibTeX users should specify bibliography style 'splncs04'.
% References will then be sorted and formatted in the correct style.
%
%	\bibliographystyle{abbrv}
%	\bibliography{conceptLattice,fuzzyLP,RoughSet}

\end{document}